\documentclass[12pt,final]{iopart}

\usepackage[english]{babel}
\usepackage[utf8]{inputenc}
\usepackage{bbold}

\usepackage{comment}
\usepackage{iopams}
\usepackage{dsfont} 
\usepackage{color}
\usepackage{graphicx}
\usepackage{float}
\usepackage{caption}
\usepackage[caption=false]{subfig}
\graphicspath{{./Images/}} 

\usepackage{braket}
\usepackage[toc,page]{appendix}

\begin{document}

\title{The average number of distinct sites visited by a random walker on random graphs}
 \author{Caterina De Bacco$^{1}$, Satya N. Majumdar$^{1}$ and Peter Sollich$^2$}
\address{$^1$ LPTMS, Centre National de la Recherche Scientifique et Universit\'e Paris-Sud 11, 91405 Orsay Cedex, France.}
 \address{$^2$ King's College London, Department of Mathematics, Strand, London WC2R 2LS, U.K.}

\ead{caterina.de-bacco@lptms.u-psud.fr}

\begin{center}
\begin{abstract}
We study the linear large-$n$ behavior of the average number of distinct sites $S(n)$ visited by a random walker after $n$ steps on a large random graph. An expression for the graph topology--dependent prefactor $B$ in $S(n)=Bn$ is proposed. We use generating function techniques to relate this prefactor to the graph adjacency matrix and then devise message-passing equations to calculate its value. Numerical simulations are performed to evaluate the agreement between the message passing predictions and random walk simulations on random graphs. Scaling with system size and average graph connectivity are also analysed.
\end{abstract}

\maketitle
\end{center}

\section{Introduction.}\label{intro}

The average number of distinct sites $S(n)$ visited by a random walker of $n$ steps moving on a graph provides important information about the geometry of the coverage of vertices on the graph. The problem of characterizing this quantity $S(n)$ as a function of time $n$ finds  interdisciplinary applications such as in target decay \cite{target} and trapping problems \cite{trapping} in chemical reactions, in the problem of annealing of point defects in crystals \cite{beeler}, in relaxation problems in disordered systems \cite{relaxation} or in problems of dynamics on the internet \cite{cattuto,networking}. Further studies have characterized the same quantity when multiple walkers are moving together \cite{Larralde,Majumdar}. \\
The problem has been widely studied (in the limit $n \gg 1$) in the case of $d$-dimensional lattices \cite{vineyard,montroll,erdosdistinct} where a number of independent studies all show that for $d>3$ the average number of distinct visited sites grows linearly in time as $S(n) ={ n}/{W(d)}$ with a prefactor $1/W(d)$ dependent on the dimension; whereas in $d=1,2$ this growth is slower, with $S(n) =\sqrt{{8n}/{\pi} }$ and $S(n)={\pi n}/{\ln n}$, respectively. In the case of Bethe lattices of connectivity $k$ the behaviour is linear again \cite{RWbethe}, with a prefactor dependent on the lattice connectivity $S(n) = [({k-2})/({k-1})]n $. 
This problem has been tackled also in the cases of graphs different from lattices or random graphs by using the spectral dimension $\tilde{d}$. Under certain assumptions $S_{i}(t) \sim t^{\min\{1,\tilde{d}\}}$ for $t \rightarrow \infty$ and $\tilde{d}\neq 2$. The quantity $\tilde{d}$ has been calculated for complex types of graphs such as decimable fractals, bundled structures, fractal trees and  $d$-simplex. See \cite{burioni2005,redner2001} for an overview. Nonetheless the determination of the prefactor remains an open questions for these complex types of graphs.

The situation where the underlying topology is a random network has only recently been studied; in particular it has been found that for Scale-Free graphs (SF) \cite{WS,SF} (in the time regime $n\gg1$) one recovers the linear behaviour $S(n) \sim n$ seen in both Bethe lattices and $d$-dimensional lattices for $d\geq 3$. However, there is very limited information on the prefactor $B$ describing this linear behavior $S(n)=Bn$ on random networks. Indeed all the studies referred to above are based on a scaling ansatz and on the analysis of numerical simulations; neither provides a theoretical framework that fully characterizes the prefactor $B$ to the same extent as has been achieved for lattices. The difficulty in setting up a theoretical model to characterize this prefactor is due to the asymmetry between forward and backward steps during the walk; this asymmetry is induced by the random nature of the graph structure, where nodes have a number of neighbours (degree) that is a random quantity extracted from a probability distribution.  \\
In this work we combine a general generating function approach, valid also for lattices, with the cavity formalism \cite{sg87,rogers2008} that has proved to be useful in a wide range of other problems in statistical physics \cite{ipc}. We derive an approximate expression for the topology dependent prefactor $B$ that is valid in the thermodynamic limit of large graphs, and for $n \gg 1$. We develop message-passing equations to calculate its value and perform numerical simulations on different graph topologies. Finally we describe the behaviour of $S(n)$ in three different time regimes through scaling considerations. 
We propose this framework as an alternative tool to the standard ones used in the case of lattices.  \\The paper is organized as follows: in section \ref{model}  we introduce the general model and the notation used to describe a random walk on random networks. Section \ref{Sgen} sets out the generating function approach to the problem. In section \ref{SRN} we then adapt it to the particular case of random networks. Our main results are derived using message-passing techniques in section \ref{MPeq}, leading to an explicit relation between the topology dependent prefactor and the cavity marginals. In section \ref{sim} we present and discuss the results of numerical simulations, including the scaling for finite graphs. We conclude in section \ref{conclusions} with a brief summary and outlook.
\section{Random walks on graphs.}\label{model}

Given a random graph $\mathcal{G}(\mathcal{V},\mathcal{E})$ with $V=|\mathcal{V}|$ nodes and $E=|\mathcal{E}|$ edges,
we denote the neighbourhood of a node $i \in \mathcal{V}$ by $\partial i$, and its degree, i.e.\ the number of neighbours, by $k_i = |\partial i |$.
An overall characterization of the graph topology is then provided by the distribution of the degrees $k_i$, which we write as $P(k)$. 
\\
Introducing matrix notation we define the graph adjacency matrix $A$ as the matrix with entries 
\begin{equation}
a_{ij} = \left\{
  \begin{array}{l l}
    1 & \quad \mbox{if $(i,j) \in \mathcal{E}$}\\
    0 & \quad \mbox{otherwise}
  \end{array} \right.
  \end{equation}
The nonzero entries of $A$ then indicate which pairs of nodes are connected by an edge. We do not consider self-loops, thus $a_{ii}=0$. Throughout we will assume that the graph is singly connected. Should the original random graph have disconnected pieces, we discard all except for the largest connected component. \\
A random walk on a graph is a path $\gamma=\{v_{0}, v_{1}, \dots, v_{n} \}$ made up of successive random steps between adjacent nodes $v_{i}$ on the graph, starting from a given node $v_{0} \in \mathcal{V}$. Steps are performed according to a transition probability from a node $i$ to an adjacent node $j$ given by:
\begin{equation}
w_{ij}=\frac{a_{ij}}{k_{i}} 
\end{equation}
All adjacent neighbours of $i$ then have equal probability of being reached in a step starting from $i$. In matrix notation we define the transition matrix $W$ as the matrix with entries $w_{ij}$. Defining also  $D$ as the diagonal matrix with entries $\delta_{ij}k_i$, we have the relation:
 \begin{equation}\label{transfermatrix}
  W=D^{-1}A
\end{equation}  
We denote the probability of reaching node $j$ in $n$ steps starting from node $i$ as $G_{ij}(n)$.\\
 With these definitions, given an $n$-step random walk $\gamma=\{v_{0},v_{1}, \dots, v_{n} \}$, the probability of reaching node $v_{n}$ starting from node $v_{0}$ along this path is the product:
 \begin{equation}
  \prod_{i=0,\dots,n-1} {\frac{1}{k_{i}} }= \frac{1}{k_{0}}\,\frac{1}{k_{1}}\times \dots \times \frac{1}{k_{n-1}}
  \end{equation}
   In general, in order to compute $G_{ij}(n)$ one has to consider all possible random walks connecting $i$ to $j$ in $n$ steps. Using the transition matrix $W$ we can write this probability as:
   \begin{equation}\label{GW}
   G_{ij}(n)=[W^{n}]_{ij} =[(D^{-1}A)^{n}]_{ij}\qquad.
   \end{equation}

\section{Average number of distinct sites: general results.}\label{Sgen}
We are interested in finding the average number of distinct sites $S_{i}(n)$ visited by a random walker taking $n$ steps on a graph starting at node $i$.\\
In this section we derive general results that are valid for any graph topology, including in particular the case of $d$-dimensional lattices. We use the formalism of generating functions, a tool that has been used to calculate $S_{i}(n)$ on lattices \cite{RWbethe,montroll} as well as other quantities of interest in the study of random walks on networks \cite{burioni2005,noh,target,fisher1984}. 
We denote by $F_{ij}(n)$ the probability of reaching site $j$ for the first time  after $n$ steps for a random walk starting at site $i$; note that for the case $i=j$ we define ``reaching'' as ``returning to'' so that $F_{ii}(0)=0$. We also define $H_{ij}(n)$ as the probability that site $j$ has been visited at least once in $n$ steps by a random walker starting at site $i$, and let $q_{j}(n)$ be the probability that a walker starting at site $j$ does not return to it within $n$ time steps.

With these definitions the average number of distinct sites visited by time $n$ (i.e.\ after $n$ steps), starting at node $i$, can be written as:
\begin{eqnarray}
S_{i}(n)&=&
\sum_{j \in \mathcal{V}} H_{ij}(n) \label{SH}
\end{eqnarray}

{{Now if a node $j$ has been visited at least once in a walk of $n$ steps starting at node $i$, we can call the time of the final visit of the walk $m\leq n$ and by definition the walk then never returns to $j$ in the remaining $n-m$ steps.

Thus we can write the convolution:
\begin{equation}\label{HGq}
H_{ij}(n)=\sum_{m=0}^{n}G_{ij}(m)q_{j}(n-m)
\end{equation}
The generating function (or $z$-transform) of a quantity $f(n)$ is defined as $\hat{f}(z)=\sum_{n=0}^{\infty}z^{n}f(n)$, with $z \in [0,1)$, and has
the property that the $z$-transform of a convolution is the product of the $z$-transforms. The $z$-transform of (\ref{HGq}) is then
\begin{equation}\label{HGztr}
\hat{H}_{ij}(z)=\hat{G}_{ij}(z)\hat{q}_{j}(z) 
\end{equation}
We now want to write everything in terms of $\hat{G}_{ij}(z)$ and so need to find a relation linking $\hat{q}_{j}(z)$ to $\hat{G}_{ij}(z)$, which we do via the first passage time probability $F_{jj}(n)$. The probability of returning to node $j$  for the first time after exactly $n$ steps can be written as:
\begin{equation}
q_{j}(n-1)-q_{j}(n)=F_{jj}(n)
\end{equation}
Taking the $z$-transform of this expression and noting that $q_{j}(0)=1$, $\hat{q}_{j}(z)=\sum_{n=0}^{\infty}z^{n}q_{j}(n)$ and $\hat{F}_{jj}(z)=\sum_{n=1}^{\infty}z^{n}F_{jj}(n)$ we have:
\begin{eqnarray}
z\sum_{n=1}^{\infty}q_{j}(n-1)z^{n-1} - \sum_{n=1}^{\infty}q_{j}(n)z^{n}&=&\sum_{n=1}^{\infty}F_{jj}(n)z^{n}\\
z\hat{q}_{j}(z) -[\hat{q}_{}(z)-1]=1-(1-z)\hat{q}_{j}(z)&=&\hat{F}_{jj}(z)
\end{eqnarray}
Hence:
\begin{equation}\label{qtr}
\hat{q}_{j}(z)=\frac{1-\hat{F}_{jj}(z)}{1-z}
\end{equation}
We now relate the generator $G_{jj}(n)$ to the first passage time probability $F_{jj}(n)$. The probability of arriving at node $j$ in $n$ steps starting at the same node $j$, can be seen as the sum of the probabilities grouped according to how often $j$ is visited overall: we can reach $j$ for the first time after $n$ steps; or a first time at $n_{1}<n$ and a second time after another $n-n_{1}$ steps; or a first time at $n_{1}<n$, a second time after another $n_{2}-n_{1}$ steps and a third time after a final $n-n_{2}$ steps, and so on. Mathematically this can be written as:
\begin{equation} \hskip -2.5cm
G_{jj}(n)=F_{jj}(n)+\sum_{n_{1}=0}^{n}F_{jj}(n_{1})F_{jj}(n-n_{1})+ \sum_{n_{2}=0}^{n}\sum_{n_{1}=0}^{n_{2}}F_{jj}(n_{1})F_{jj}(n_{2}-n_{1})F_{jj}(n-n_{2})+ \dots \hskip 0.1cm
\end{equation}
To make the convolution structure clearer, we have included the extreme values (e.g.\ $n_{1}=0$ and $n_1=n$ in the first sum) here even though -- because $F_{jj}(0)=0$ -- they do not contribute. Taking the $z$-transform of both sides one sees that
\begin{equation}
\hat{G}_{jj}(z)=1+\hat{F}_{jj}(z)+\hat{F}_{jj}^{2}(z)+ \dots=\frac{1}{1-\hat{F}_{jj}(z)}
\end{equation}
Substituting this result into (\ref{HGztr}) using (\ref{qtr}) we obtain:
\begin{equation}
\hat{H}_{ij}(z)=\hat{G}_{ij}(z)\frac{1-\hat{F}_{jj}(z)}{1-z}=\frac{1}{(1-z)}\frac{\hat{G}_{ij}(z)}{\hat{G}_{jj}(z)}
\end{equation}
This can now be inserted into (\ref{SH}) to give finally the $z$-transform of the average number of distinct sites visited starting from site $i$:
\begin{equation}\label{SG}
 \hat{S}_{i}(z)=\frac{1}{1-z}\sum_{j \in \mathcal{V}}\left[  \frac{\hat{G}_{ij}(z)}{\hat{G}_{jj}(z)} \right]
\end{equation}
One sees that the underlying quantity of central interest for our problem is $\hat{G}_{ij}(z)$. The result of equation (\ref{SG}) is valid in general, i.e.\ regardless of the graph topology. We note that to understand the large $n$-behaviour of $S_i(n)$ we need to consider $\hat{S}_{i}(z)$ near $z=1$. Specifically, if as expected for $V\to\infty$ we have $S_i(n)=Bn$ for large $n$, then the $z$-transform will diverge for $z\to 1$ as $\hat{S}_i(z)=B/(1-z)^2$. To calculate $B$ we thus need to understand the behaviour of $\hat{G}_{ij}(z)$ for $z\to 1$.} 

\section{Average number of distinct sites: random graph results.}\label{SRN}
In this section we will derive an expression for $G(n)$, the matrix with entries $G_{ij}(n)$, where the dependence on the graph size for large graphs is explicit. Here we will for the first time have to restrict the type of graph: as explained below, we require that the eigenvalue spectrum of $A$ has a nonzero gap.

As we saw in section \ref{model}, in the case of random graphs we have $ G(n)=W^{n} =(D^{-1}A)^{n}$ and hence $\hat{G}(z)=(\mathds{1} -zD^{-1}A)^{-1}$, which relates the propagator $G$ to the graph topology via the adjacency matrix $A$.

To transform to a symmetric matrix whose properties are simpler to understand, we rewrite this as
\begin{eqnarray}\label{GR}
\hat{G}(z)
&=&D^{-1/2}\hat{R}(z)D^{+1/2}
\end{eqnarray}
in terms of the matrix
\begin{equation}\label{R}
\hat{R}(z)=(\mathds{1} - zD^{-1/2}AD^{-1/2})^{-1}
\end{equation}
This matrix is now clearly symmetric, and we can diagonalize it as
\begin{equation}
\hat{R}=P{\bLambda}P^{T}
\end{equation}
where the matrix $P$ has as columns the eigenvectors of $\hat{R}$ and ${\bLambda}$ is a matrix containing the eigenvalues of $\hat{R}$ on the diagonal.\\
In terms of the normalized adjacency matrix $M=D^{-1/2}AD^{-1/2}$ \cite{spectral}, one has
\begin{equation}\label{RM}
\hat{R}(z)=(\mathds{1}-zM)^{-1}
\end{equation}
In the following we use Dirac bra-ket notation \cite{braket} to denote the eigenvectors $\ket{u_{k}}$ of $M$. If $\ket{u_{k}}$ is one such eigenvector and $\lambda_{k}$ the corresponding eigenvalue, then 
\begin{equation}
M\ket{u_{k}}=\lambda_{k}\ket{u_{k}}
\end{equation}
and it follows that
\begin{eqnarray}
\hat{R}(z)\ket{u_{k}}&=&
(1-z\lambda_{k})^{-1}\ket{u_{k}}
\end{eqnarray}
In words, $\hat{R}(z)$ has the same eigenvectors $\ket{u_{k}}$ as $M$ but with corresponding eigenvalues $1/(1-z\lambda_{k})$. \\
From spectral graph theory \cite{spectral} we know that the $z$-independent matrix $M$ has eigenvalues all lying in the range $[-1,1]$.

By direct substitution into the eigenvalue equation for $M$ one sees that the vector with entries $u_{1,i}=c\sqrt{k_{i}}$ is an eigenvector with eigenvalue $\lambda_{1}=1$. The constant $c$ is found from the normalization condition $\braket{u_{1}|u_{1}}=\sum_{i=1}^{V}{u^{2}_{1,i}}=1$ as
$c^{-1}=\sqrt{V\,\langle k \rangle }$ where $\langle k \rangle=\sum_{j
  \in V} k_{j} /V$ is the average degree of the graph. If the graph is singly connected then there are no other eigenvectors with eigenvalue 1, so we can order the eigenvalues as
\begin{equation}\label{lambda}
1=\lambda_{1}>\lambda_{2}\geq \ldots \geq \lambda_{V} \geq -1
\end{equation}
(The fact that the eigenvalues lie between $-1$ and $1$ can also be seen from the Perron-Frobenius theorem \cite{frobenius,perron}, given that the entries of $\ket{u_{1}}$ are all positive and $\lambda_{1}=1$.)

Splitting off the contribution from $\lambda_1$, we can now write the eigenvector decomposition of $\hat{R}(z)$ as
\begin{eqnarray}
\hat{R}(z)
&=& \ket{u_{1}} \bra{u_{1}} \frac{1}{1-z} + \sum_{k=2}^{V}  \ket{u_{k}} \bra{u_{k}} \frac{1}{1-z\lambda_{k}}
\end{eqnarray}
and clearly the first term will be dominant in the limit $z \rightarrow 1$ that we need to consider. \\
With the shorthand
\begin{eqnarray}
C(z)&=& \sum_{k=2}^{V}  \ket{u_{k}} \bra{u_{k}} \frac{1}{1-z\lambda_{k}} \label{B}
\end{eqnarray}
for the second term, we can then write
\begin{eqnarray}
\hat{R}_{ij}(z)&=& \frac{\sqrt{k_{i}k_{j}}}{V \langle k \rangle} \frac{1}{1-z} + C_{ij}(z) \label{RC}
\end{eqnarray}
From equation (\ref{GR}) we have 
$\hat{G}_{ij}(z)=({k_{j}}/{k_{i}})^{1/2} \hat{R}_{ij}(z)$, so the analogous representation for $\hat{G}(z)$ reads
\begin{eqnarray}
\hat{G}_{ij}(z)&=& \frac{k_{j}}{V \langle k \rangle}\frac{1}{1-z} +\sqrt{\frac{k_{j}}{k_{i}}}C_{ij}(z)
\end{eqnarray}
We can now substitute these expressions into equation (\ref{SG}) to obtain:
\begin{equation}
\hat{S}_{i}(z)= \frac{1}{1-z} \sum_{j \in V} \left\{  \frac{  k_{j}}{ \hat{R}_{jj}(z)V \langle k \rangle (1-z)}  + \frac{\sqrt{\frac{k_{j}}{k_{i}}}C_{ij}(z)V \langle k \rangle (1-z)}{ k_{j} + C_{jj}(z)V \langle k \rangle (1-z)} \right\}   \label{eqS}
\end{equation}

In the following we will consider first the limit $V \rightarrow \infty$ and then the limit $z \rightarrow 1$. This order of taking the two limits is important to get physical results, as we explain in more detail below. Note that the denominators in the two terms of (\ref{eqS}) are identical but written in two different forms that will make the limit procedure clearer.

The large $V$-limit is simple to take in (\ref{RC}), giving $\lim_{V \rightarrow \infty} \hat{R}_{jj}(z)=C_{jj}(z)$. We are assuming implicitly here that $C(z)$ has a well-defined limit for $V\to\infty$. This requires in particular that $\lambda_2$ stays away from 1, i.e.\ that the spectrum of $M$ has a nonzero gap $1-\lambda_2$ between the leading and first subleading eigenvalue for $V\to\infty$. This is generally true for regular \cite{friedman1991,broder1987}, ER \cite{farkas2001,furedi81} and scale-free \cite{chung2003,farkas2001} random graphs, but not for lattices, where the eigenvectors are Fourier modes whose eigenvalue approaches 1 smoothly in the large wavelength (zero wavevector) limit.

In the second term of (\ref{eqS}), the first term in the denominator can be neglected for $V\to\infty$ at fixed $z<1$, giving
 \begin{equation}
\lim_{V \rightarrow \infty} \hat{S}_{i}(z)=  \frac{1}{1-z}\sum_{j \in V}\left\{  \frac{k_{j}}{C_{jj}(z)V \langle k \rangle (1-z)} + \frac{\sqrt{k_{j}}C_{ij}(z)}{\sqrt{k_{i}}C_{jj}(z)} \right\}
\end{equation}
Now we take the limit $z \rightarrow 1$, in which the second term becomes negligible compared to the first. With the assumption of a nonzero gap, $C_{jj}(z)$ also has a finite limit for $z\to 1$ so that we can define
  \begin{equation}
  \lim_{z \rightarrow 1} \left[ \lim_{V \rightarrow \infty}\hat{R}_{jj}(z)\right]=  \lim_{z \rightarrow 1} C_{jj}(z)=R_{j}\label{limR}
   \end{equation}
   and get finally
 \begin{equation}\label{limS}
\lim_{V \rightarrow \infty}\hat{S}_{i}(z)= \frac{1}{V \langle k \rangle (1-z)^{2}} \sum_{j \in V} \frac{k_{j}}{R_{j}}
\end{equation}

as the asymptotic behaviour for $z\to 1$.

This has exactly the $1/(1-z)^2$ divergence we were expecting, 
and gives us the prefactor of the large $n$-asymptote of the number of distinct sites visited:
 \begin{equation} \label{Sn}
 \lim_{V \rightarrow \infty} S_{i}(n)=B\,n
\end{equation}
where
 \begin{equation} \label{pref}
B= \frac{1}{V \langle k \rangle} \sum_{j \in V} \frac{k_{j}}{R_{j}}
\end{equation}
We can make three observations.
Firstly, if we had inverted the order of taking the limits and fixed $V$ while taking $z\to 1$, then we would have had $\hat{R}_{jj}(z)= k_j/[V\langle k\rangle (1-z)]$ to leading order. The second term in (\ref{eqS}) would have disappeared in the limit, so that {{
 \begin{eqnarray}
\hat{S}_{i}(z) &=& \frac{1}{1-z} \sum_{j \in V} \frac{  k_{j}}{ \hat{R}_{jj}(z)V \langle k \rangle (1-z)}  
=\frac{1}{1-z} V
\end{eqnarray}
}

to leading order near $z=1$. This $1/(1-z)$ divergence of $\hat{S}_{i}(z) $ implies $\lim_{ n \rightarrow \infty} S_{i}(n)=V$, a result which is clear intuitively: if we keep the graph size finite then in the limit of large times the random walk will cover the entire graph, i.e.\ visit all nodes at least once.\\
Secondly, from equation (\ref{limR}) we can see that the information one needs to calculate $B$ resides in the quantities
$C_{jj}(z)=\sum_{k=2}^{V} u^{2}_{k,j}/(1-z\lambda_{k})$, where the $u_{k,j}$ are the components of the eigenvectors $\ket{u_{k}}$ of $M$ and the $\lambda_k$ the eigenvalues. So knowing the full spectrum of $M$ and the associated eigenvector statistics would in principle solve our problem of determining $B$. While this is feasible computationally for finite and not too large $V$, we are not aware of a method that would work in the thermodynamic limit $V\to\infty$.\\
Thirdly, although the index $i$ appears on the left hand side of equation (\ref{Sn}), representing the initial node of the walk, it does not appear on the right. This means that the average number of distinct sites visited in the large $n$ limit does not depend on the starting node, and therefore we can drop the index $i$ from the left hand side of (\ref{Sn}). In particular, even for graphs with broad degree distributions such as scale-free graphs, the number of distinct sites visited will be the same whether we start the walk from a hub (a node with high degree) or a dangling end of the graph (a node with degree one) -- provided of course $n$ is large enough.

\section{The message-passing equations.}\label{MPeq}
From expression (\ref{pref}) we see that, for a given graph, we need
to calculate the quantity $\frac{k_{j}}{R_{j}}$. Although we
know the entries of the inverse
$\hat{R}^{-1}_{ij}(z)=\delta_{ij}-z {a_{ij}}(k_ik_j)^{-1/2}$, it is not
straightforward to characterize $\hat{R}_{jj}(z)$. We could find the value $R_{j}$ either by calculating $\lim_{z \rightarrow 1}C_{jj}(z)$ where $C_{jj}(z)=\sum_{k=2}^Vu_{k,j}^2/(1-z\lambda_k)$ or by directly inverting the matrix $\hat{R}^{-1}(z)=[\mathds{1} - zD^{-1/2}AD^{-1/2}]$. Unfortunately both of these two methods are prohibitive computationally, already for individual graphs of large size $V$ and even more so if in addition we want to average the results over an ensemble of random graphs.

Our aim, then, is to find a viable alternative method that will allow us to characterize the value of $\hat{R}_{jj}(z)$, and thus calculate $\lim_{n\rightarrow \infty}S(n)$ through (\ref{Sn}) and (\ref{pref}). We draw for this on methods that have been deployed in the calculation of sparse random matrix spectra \cite{rogers2008}. That a connection to spectral problems should exist is suggested by the fact that $z\hat{R}(z)=(z^{-1}\mathds{1} - D^{-1/2}AD^{-1/2})^{-1}$: up to a trivial rescaling, $\hat{R}(z)$ has the structure of a resolvent (with parameter $z^{-1}$) for the random matrix $D^{-1/2}AD^{-1/2}$, and it is from such resolvents that spectral information is normally derived, in an approach that in the statistical physics literature goes back to at least Edwards and Jones \cite{edwards1976}. Accordingly the two steps we will need to take mirror closely those used to find resolvents of sparse random matrices in  \cite{rogers2008}: we first write the $\hat{R}_{jj}(z)$ as variances in a Gaussian distribution with covariance matrix $\hat{R}^{-1}(z)$, and then exploit the fact that this distribution has a graphical model structure to derive cavity equations from which these variances can be found.

\subsection{Multivariate Gaussian representation.}

The first step is simple: we define a vector of random variables $(x_{1},\dots,x_{V})$ and assign to this the zero mean Gaussian distribution
\begin{eqnarray}
P(\bar{x})&\propto& e^{-\bar{x}^{\rm T}\hat{R}^{-1}(z)\bar{x}/2 }
= e^{-\bar{x}^{T}(\mathds{1}-zD^{-1/2}AD^{-1/2})\bar{x}/2}\label{multilap}
\end{eqnarray}
The marginal distribution of any component of the vector, obtained by integrating $P(\bar{x})$ over all other components, is then also Gaussian:

\begin{equation}\label{marginalp}
P(x_{j}) \propto e^{-x^{2}_{j}/(2{v_{j}})} 
\end{equation}
with variance $v_{j}=\langle x_{j}^{2}\rangle=\hat{R}_{jj}(z)$. Our goal is now to calculate these marginal variances efficiently, i.e.\ without a full matrix inversion.


The key property of the probability distribution (\ref{multilap}) is that it can be written in the form
\begin{eqnarray}
P(\bar{x})
&=& \prod_{i\in \mathcal{V}}e^{-x_{i}^{2}/2}\prod_{(ij)\in \mathcal{E}}e^{z x_{i}x_{j}
(k_{i}k_{j})^{-1/2}} \label{jointd}
\end{eqnarray} 
As this factorizes into contributions associated with the nodes and edges of the underlying graph, it defines what is known as a graphical model \cite{ipc}. On such a graphical model, marginal distributions can be obtained using message-passing, or cavity, equations.

\subsection{Cavity equations.}
For completeness, we summarize briefly the derivation of the message-passing equations, also known as sum-product algorithm \cite{ipc}. We focus on trees, i.e.\ graphs that do not contain any loops, where the equations are exact, and leave for later a discussion of the extent to which they apply also to large random graphs.
Write generally $\phi_{i}(x_{i})$ for the factor in $P(\bar{x})$ associated with node $i$ and $\psi_{ij}(x_{i},x_{j})$ as the interaction term between nodes $i$ and $j$. In our case we have:

\begin{eqnarray}
\psi_{ij}(x_{i},x_{j})&=& e^{z x_{i}x_{j}(k_{i}k_{j})^{-1/2}} \label{interaction} \\
\phi_{i}(x_i) &=& e^{-x_{i}^{2}/2} \label{field}
\end{eqnarray}

To calculate the marginal distribution of $x_j$, we could imagine first removing all edge factors $\psi_{ij}(x_j,x_i)$ from $P(\bar{x})$, where $i$ runs over all neighbours of $j$. The tree is now split into subtrees rooted at each neighbour $i$, and one can define the cavity marginal of $i$, $\nu_{i \rightarrow j}(x_{i})$ as the marginal that is obtained from a (suitably renormalized) probability distribution containing only the factors from the relevant subtree. To get the marginal of $x_j$, we now just need to reinstate the missing edge factors as well as the node factor at $j$ and integrate over the values of the nodes that we have not yet marginalized over, namely, the neighbours $i$: 
 \begin{equation}\label{marginalprod_gen}
 P(x_{j}) \propto \phi_{j}(x_j) \prod_{i\in \partial j}\int
dx_{i}\, \psi_{ji}(x_j,x_i) \nu_{i \rightarrow j}(x_{i})
 \end{equation} 
One can call the quantities $\nu_{i \rightarrow j}(x_{i})$ messages sent from $i$ to $j$, or cavity marginals: each message tells node $j$ what the marginal of its neighbour $i$ would have been if the edge between them had been severed.

The cavity marginals can now be obtained from an analogous relation. To get $\nu_{i \rightarrow j}(x_{i})$, one can think of removing all edges connecting $i$ to its neighbours $l$ other than $j$; note that the edge connecting $i$ to $j$ has already been taken out in the definition of the cavity marginal. This generates independent subtrees rooted at the neighbours $l$, and the marginals at these nodes are $\nu_{l \rightarrow i}(x_{l})$. Reinstating removed edge factors and marginalizing over neighbours then yields
 \begin{equation}\label{cavitymess_gen}
\nu_{i \rightarrow j}(x_i) \propto \phi_{i}(x_i) \prod_{l\in \partial i \setminus j}\int
dx_{l}\, \psi_{il}(x_i,x_l)\nu_{l \rightarrow i}(x_{l})
 \end{equation} 
On a tree these equations can be solved by e.g.\ starting at leaf nodes, where simply $\nu_{i\rightarrow j}(x_{i})\propto \phi_{i}(x_i)$, and then sweeping through the tree in a way that calculates each message once messages have been received from all neighbours except the intended recipient of the message. Note that two messages are needed per edge, one in each direction. Once all messages have been found, the marginals can be deduced from (\ref{marginalprod_gen}).

On graphs with loops, the message-passing equations (\ref{marginalprod_gen}) and (\ref{cavitymess_gen}) are no longer exact: when we remove all edges around node, its neighbours may then still be correlated because of loops, and we cannot factorize their joint distribution into a product of cavity marginals.
The cavity method, also known as Bethe-Peierls approximation \cite{ipc}, consists in neglecting such correlations. The set of equations  (\ref{cavitymess_gen}) for the cavity marginals is then viewed as a set of fixed point equations that typically have to be iterated to convergence (see below). Clearly the marginals we deduce in the end are approximate. Nevertheless the method remains useful for us because we expect the approximation to become exact for random graphs in the limit of large $V$. The reason is that typical loop lengths diverge (logarithmically) with $V$, so that the graphs become locally tree-like \cite{ipc,wormald1999}. The correlations that the cavity method ignores then weaken as $V$ grows, making the approach exact for large $V$.

Specializing now to our Gaussian graphical model, the cavity marginals must also be Gaussian and we can write them as

\begin{equation}\label{marginalvariance}
\nu_{l \rightarrow i}(x_{l}) \propto e^{-x_{l}^{2}/(2 v_{l}^{(i)})}
\end{equation}
which defines the cavity variances $v_{l}^{(i)}$.
%
%
Inserting (\ref{field}) and (\ref{interaction}) into the general message passing equation (\ref{cavitymess_gen}) and carrying out the resulting Gaussian integrals gives then
\begin{eqnarray}\label{cavityvariance}
v_{i}^{(j)}&=& k_{i} \left(k_{i} -z^{2}\sum_{l \in \partial i\setminus j} \frac{v_{l}^{(i)}}{k_{l}} \right)^{-1} 
\end{eqnarray}
%
while for the full marginals one obtains analogously
\begin{equation}\label{marginals}
v_{j}=k_{j}\left( k_{j} -z^{2}\sum_{i \in \partial j} \frac{v_{i}^{(j)}}{k_{i}} \right)^{-1}
\end{equation}
These two relations are the direct analogues of Eqs.\ (11) and (12) in \cite{rogers2008}.

The variances $v_{j}$, when calculated in the limit ${z\rightarrow1}$, are the quantity of interest for our problem as $v_{j}=\langle x_{j}^{2}\rangle=R_{j}$. They are known once the cavity variances have been obtained by solving (\ref{cavityvariance}).

In practice we use the rescaled cavity variances
\begin{equation}\label{mij}
m_{i\rightarrow j}=\frac{v_{i}^{(j)}}{k_{i}}
\end{equation}
as messages from node $i$ to node $j$. With this definition and using (\ref{cavityvariance}) for ${z\rightarrow 1}$ the cavity equations are:
\begin{equation}
\label{cavequ_fp}
m_{i\rightarrow j}=\left(k_{i} -\sum_{l \in \partial i\setminus j} m_{l \rightarrow i}\right)^{-1} 
\end{equation}

We solve these by iteration according to

\begin{eqnarray}\label{cavequ}
m_{i\rightarrow j}^{(t+1)}&=& \left( k_{i} -\sum_{l \in \partial i\setminus j}m_{l \rightarrow i}^{(t)}\right)^{-1} 
\end{eqnarray}
where $t$ represents a discrete iteration time step.

Starting from a given graph $\mathcal{G}$, a suitably chosen convergence criterion and a maximum iteration time $T_{\rm max}$, the algorithm then works as following: 
\begin{enumerate}
\item Initialize the messages $m_{i\rightarrow j}^{(0)}$ randomly.
\item Run through all edges $(ij)$ and find for each the updated messages $m_{i\rightarrow j}^{(t+1)}$, $m_{j\rightarrow i}^{(t+1)}$ from (\ref{cavequ}).
\item Increase $t$ by one.
\item Repeat steps 2 and 3 until either convergence is reached or $t=T_{\rm max}$.
\end{enumerate}
 If convergence is reached, i.e.\ the preset convergence criterion is satisfied, one can collect the results and calculate the variances $v_{j}$ using (\ref{marginals}) and (\ref{mij}):
  \begin{equation}\label{cavmar}
v_{j}=k_{j} \left( k_{j} -\sum_{i \in \partial j}m_{i\rightarrow j}\right)^{-1}
\end{equation}
where $m_{i\rightarrow j}$ are the converged messages.


If we identify $v_{j}=\langle x_{j}^{2}\rangle=R_{j}$
%
%
we can then also express directly the prefactor (\ref{pref}) in the linear asymptote in the number of distinct sites visited, $S(n)=Bn$, as
\begin{eqnarray}
B&=&\frac{1}{V \langle k \rangle} \sum_{j \in V}\frac{k_{j}}{v_{j}} \label{BMPv}\\
&=&\frac{1}{V \langle k \rangle} \sum_{j \in V} \left(k_{j} -\sum_{i \in \partial j} m_{i\rightarrow j}\right) \label{BMP}
\end{eqnarray}

There is one subtlety here that we have glossed over: the variances $v_j$ are the full marginal variances $\hat{R}_{jj}(z)$, which from (\ref{RC}) have the form $k_j/[V\langle k\rangle (1-z)]+C_{jj}(z)$. In the calculation of $B$ we need $R_{j}=\lim_{z\to 1} C_{jj}(z)$, where the contribution $\propto (1-z)^{-1}$ has been removed. Where we have taken the limit $z\to 1$ above, we therefore implicitly mean that $1-z$ needs to lie in the range $1/V \ll 1-z \ll 1$ where the divergent contribution to $\hat{R}_{jj}(z)$ is still small enough to be neglected compared to $C_{jj}(z)$. That it is then allowable nevertheless to set $z=1$ directly in the cavity equations that we solve is something that has to be checked numerically: we do indeed always find finite marginals $v_j$ from converged solutions for the cavity marginals. The divergent solution also exists as a separate fixed point, namely the trivial solution $m_{i\rightarrow j} \equiv 1$ of (\ref{cavequ_fp}), but is not accessed in our iterative solution method.

\subsection{Regular graph case.}
Before going on to numerical results for more general random graph ensembles, we briefly use the expression for the topology dependent prefactor (\ref{BMP}) to consider the particular case of a regular graph, i.e.\ a graph where $\forall i \in \mathcal{V}$ we have $k_{i}=k$. In the infinite graph size limit the graph is then effectively (up to negligible long loops) a regular tree, where each node is equivalent to all others. The quantities of interest in (\ref{cavequ_fp}), (\ref{cavmar}) must then be the same $\forall i \in \mathcal{V}$: we can write $k_{i}=k$, $v_{i}^{(j)}=v^{(j)}$, $m_{i \rightarrow j}=m$ and $v_{j}=v$. The fixed point cavity equations (\ref{cavequ_fp}) thus reduce to:
\begin{eqnarray}
m&=&  \left( k -\sum_{l \in \partial i\setminus j} m \right)^{-1}
=
[k -(k-1)m]^{-1}  \nonumber  
\end{eqnarray}
We obtain a second order equation in $m$:
\begin{equation}\label{2ndm}
m^{2}(k-1)-mk +1 =0
\end{equation}
 with solutions $m=1/(k-1)$ or $m=1$. The first solution is the one we require; the second one is the trivial solution discussed above that gives divergent variances in (\ref{cavmar}). From $m=1/({k-1})$ one can find the cavity variances and from there the full variances
\begin{equation}
v=\frac{k-1}{k-2}
\end{equation}
Substituting into the expressions (\ref{BMPv}) for the prefactor $B$ we obtain:
\begin{equation}
B=  \frac{k-2}{k-1}
\end{equation}
This result agrees with the one derived for Bethe lattices of connectivity $k$ \cite{RWbethe}. This is as expected, given that the cavity method is exact on tree graphs.

\section{Simulations.}\label{sim}
We performed numerical simulations to test the predictions from our cavity approach for the number of distinct sites visited. We used four types of graph structures: regular random graphs (Reg), Erd\H{o}s-R\'{e}nyi (ER) \cite{ER}, scale-free (SF) using a preferential attachment scheme \cite{SF} and a dedicated graph ensemble (RER) where graphs are built starting from a $k_{0}$-regular random graph, with edges then added independently with probability $p$ as in the ER  model; if $d=pV$ then the final average degree of such a graph is $\langle k \rangle =k_{0} + d$ for large $V$. This graph ensemble thus interpolates between the regular and ER cases and is similar to the one analyzed in  \cite{Monasson,WS} with the difference that here we start from a regular graph instead of a ring or a lattice. As for the preferential attachment we used the  following procedure: start with a graph of $m_{0}$ vertices and introduce sequentially $V-m_{0}$ new vertices by attaching each of them to $m$ already existing nodes. The probability to pick a certain node $i$ as one of these $m$ neighbors is proportional to its degree, $P(k_{i})\sim k_{i}$; thus high degree nodes will be more likely to be picked and hence they will increase their degree while the graph grows. These scheme leads to a power-law degree distribution $P(k) \sim k^{-\gamma}$ with $\gamma=2.9 \pm 0.1$ \cite{SF}; we empirically observe this value in our simulations. We also tried other generation methods for scale-free graphs that yield different values of $\gamma$, but as the results were qualitatively similar to those for preferential attachment we only show the latter as representative for our scale-free graph simulations.

 For each of these graph topologies we investigated three fixed sizes $V=10^{3}, 10^{4}, 10^{5}$ and different average degrees. For ER graphs we only used the giant connected component of each graph sampled, but the average degrees we consider are large enough ($\langle k \rangle \geq 4$) for this to reduce $V$ by at most by 2\%. The other types of graph have only one connected component by construction.   
For each given graph we evaluated the cavity prediction (\ref{BMP})
from a converged solution of the cavity equations (\ref{cavequ_fp}). The iterative solution using (\ref{cavequ}) converged quickly, in typically around $10$ iteration steps. We used as convergence criterion the following: convergence is reached if $\max_{(ij)\in \mathcal{E}}| m_{i\rightarrow j}^{(t+1)}-m_{i\rightarrow j}^{(t)}| < \epsilon$ for $y$ consecutive times, where we set $y=10$ and $\epsilon=10^{-5}$. The results for $B$ were averaged over $1,000$ different graph instances for $V=10^{3},10^{4}$ and $100$ instances for the bigger graphs of size $V=10^{5}$.

The cavity predictions were compared against direct simulations of unbiased random walks. Each walk starts at a randomly picked vertex and we keep track of the number of distinct visited sites as the walk progresses, with individual steps performed using the transition probabilities $w_{ij}=\frac{a_{ij}}{k_{i}}$ defined in section \ref{model}. We averaged the results over the same graph instances as used to generate the cavity predictions. Note that for each instance of a given graph type, only a single walk was performed starting from a randomly chosen initial site. 
Note that while the cavity prediction depends only on the topology of each graph, for the direct simulations there is an additional source of randomness arising from the particular random walk trajectory that is obtained on a given graph.

The issue of how the cavity predictions depend on graph size $V$ deserves a brief comment. We argued that the method should become exact in the limit $V\to \infty$, and so a priori should extrapolate our predictions for $B$ to this limit. We found, however, that for our relatively large graph sizes the predictions for different $V$ agreed within the error bars. Thus we did not perform a systematic extrapolation and simply used the predictions for $V=10^{4}$, as the largest graph size for which we could obtain a statistically large sample (1000 graph instances) of data. The fact that already $V=10^{3}$, our smallest size, is large enough to obtain results that are essentially indistinguishable from those for $V\to\infty$ is consistent with findings from cavity predictions in other contexts, see e.g. \cite{peter2013,peter2012}.
An alternative approach to evaluating the cavity predictions would have been to move from specific graph instances to solving the limiting ($V\to\infty$) integral equations for the distribution of messages across the graph. These equations  can be read off more or less directly from the cavity equations, see e.g. \cite{peter2013, sollich2014}, or obtained from replica calculations \cite{reimer2008} and then solved numerically using population dynamics. Given the good agreement between the predictions for our three different $V$ this approach would be expected to give identical predictions, so we did not pursue it.


\subsection{Simulations versus cavity predictions.}

Our first task is to verify that the cavity equations do indeed correctly predict the prefactor $B$ for random walks on large graphs.  
In figure \ref{kscaling} we plot the average number of sites $S(n)$ visited for ER graphs of degree $k=4$, 7 and 10. We plot $S(n)$ versus $Bn$, with $B$ the value taken from the cavity predictions, so that the data points should lie on the diagonal $y=x$ if the cavity predictions are accurate. We see in figure \ref{kscaling} that this is indeed the case, for graphs of size $V=10^4$. Similar levels of agreement are obtained for the other graph ensembles and sizes. The numerical data thus fully support our argument that the cavity predictions will be exact for large $V$, and show that in fact $V$ does not have to be excessively large to reach good quantitative agreement between the predictions and direct simulations.

\begin{figure}[!ht]
        \centering
          \includegraphics[width=14cm]{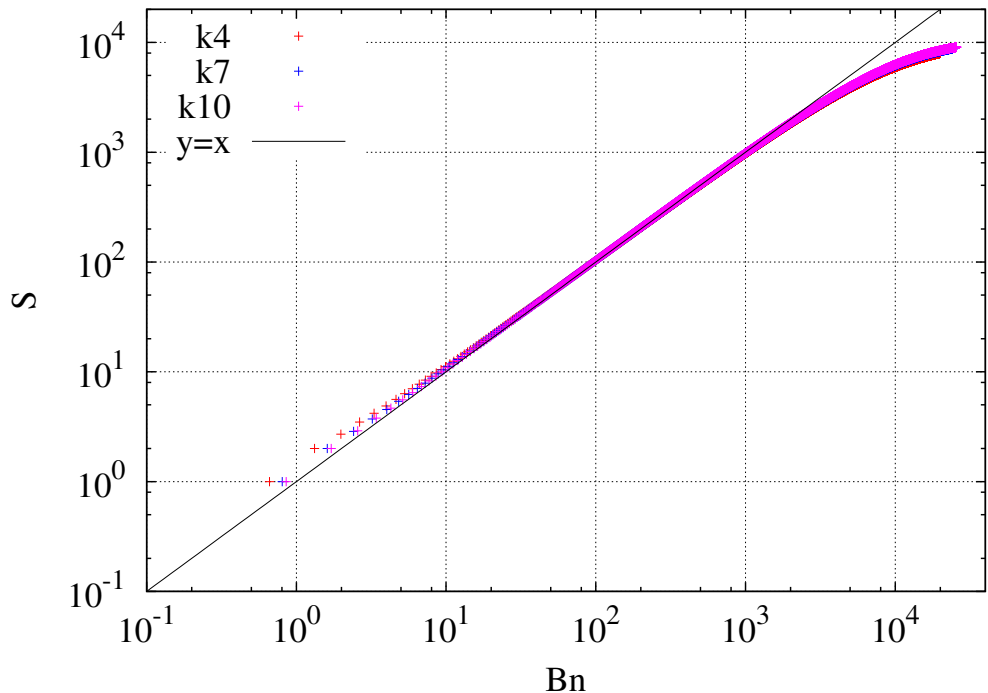}
                          \caption{Average number of distinct sites visited, $S(n)$ for random walks on ER graphs of size $V=10^{4}$. $S(n)$ is plotted against $Bn$ with the prefactor $B$ as predicted by the cavity method (\ref{BMP}), for different average degrees $\langle k\rangle=4,7,10$ as shown in the legend. In the linear regime, before the random walk starts to saturate the graph, data points lie on the diagonal, showing excellent agreement between predictions and direct simulations.}\label{kscaling}
\end{figure}

\subsection{Dependence on graph topology.}

We next look more systematically at how the prefactor $B$ in the large $n$-behaviour $S(n)=Bn$ depends on the topology of the graphs we study. 
In figure \ref{Dk} we report the dependence of the cavity prediction for $B$ on average node degree $\langle k\rangle$, for the four different graph ensembles we studied. We found that for each graph type a hyperbolic fit of the form $B(\langle k\rangle)=\frac{\langle k\rangle-c_{1}}{\langle k\rangle-c_{2}}$ gives a good description of the data, with the parameters $c_{1},c_{2}$ dependent on the graph topology but best fit values always satisfying $c_{1}=c_{2}+1$. Thus we could interpret the generic graph result as the one for a regular graph with effective degree $\langle k\rangle - c_2+1$. This is intriguing as it suggests that the effect of changing the average degree is quite similar between the different graph types.

Looking at quantitative differences between graph ensembles, we observe that the prefactor $B$ is smallest for given connectivity (average degree) when the graph is regular. Heterogeneity in the node degrees thus generically seems to {\em increase} the number of distinct sites a random walk will visit, a result that seems to us non-trivial and would be interesting to investigate as a broader conjecture: could there be a lower bound $B\geq (\langle k\rangle - 2)/(\langle k\rangle - 1)$? 
 If this were the case, one may wonder whether this is related to the spectral gap of a given graph, which is maximal for regular random graphs \cite{alon1986,nilli1991}. Indeed the impact of the gap would appear in the numerator of the prefactor (\ref{pref}) through equation (\ref{limR}) and by using the definition $C_{jj}(z)=\sum_{k=2}^{V} u^{2}_{k,j}/(1-z\lambda_{k})$. Nonetheless the gap contribution could be balanced off by the square of the eigenvector entries $u^{2}_{k,j}$ of the matrix $R$ which can be of order $O(1)$ or $O(1/V)$ depending on the eigenvector localization or delocalization respectively. For instance scale-free graphs have been shown empirically to be localized (when considering the adjacency matrix), i.e. only a few eigenvector entries are non-zero and these correspond to the high degree nodes \cite{farkas2001}, whereas for ER graph the amplitude of the eigenvalue entries is evenly distributed among all the nodes;  this difference can be detected for instance by calculating the inverse participation ratio \cite{reimer2008,farkas2001}. In order to make a more rigorous statement one would then need to consider these two aspects at the same time but the absence of a general analytical characterization for either the eigenvalues or the eigenvector entries makes this difficult. 

One could also ask whether at given $\langle k\rangle$, $B$ is always increasing with some measure of spread of degrees such as the variance $\langle k^2\rangle - \langle k\rangle^2$. For our admittedly limited choice of graph ensembles it is certainly true that the scale-free graphs (SF), which have the broadest degree distributions, also give the largest $B$. Below them are the ER graphs. The RER graphs, finally, with their character intermediate between regular and ER, also have prefactors $B$ that lie between those of the ER and regular graphs.

\begin{figure}[!ht]
        \centering
\includegraphics[width=14cm]{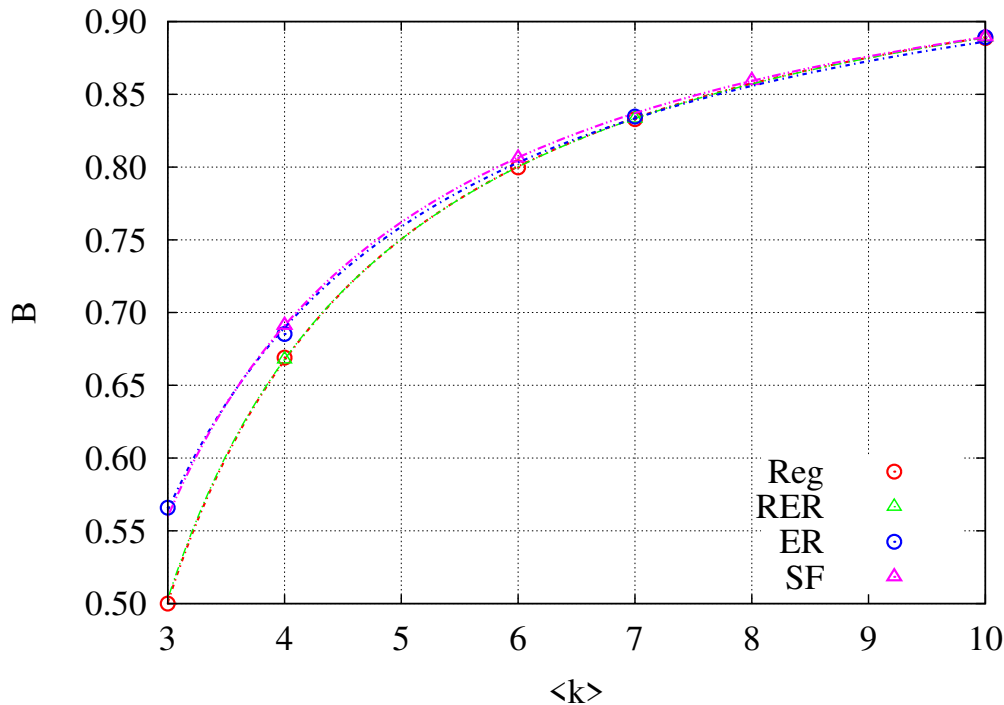}
                          \caption{Prefactor $B$ predicted by cavity method as a function of average degree, for different graph types as shown in the legend. The lines represent hyperbolic fits; see text for details. Note that the results for Reg and RER are essentially on top of each other, and the same is true for ER and SF.}   \label{Dk}
\end{figure}


\subsection{Finite-size effects and scaling.}

We can use our numerical simulation results to enquire also about finite-size effects, describing the behaviour of $S(n)$ on graphs of large but finite size $V$. Our derivation of $B$ and its prediction using cavity techniques was done taking a large $V$-limit so cannot make statements about this regime; instead we will have to rely on physical intuition to construct a suitable finite-size scaling ansatz.

From inspection of the numerical simulations, we can distinguish a number of time regimes. Initially $S(n)$ is linear in $n$ with prefactor $1$. This is greater than the large $n$ prediction $Bn$ with a prefactor $B<1$, because the random walker has not yet had much opportunity to return to previous sites; in particular one has, trivially, $S(1)=1$, ignoring the starting site $v_{0}$. 

For larger $n$ one finds the predicted linear growth with prefactor $B<1$, i.e.\ $S(n)=Bn$. Once $Bn$ becomes comparable to $V$, a crossover to sublinear growth takes place, and finally $S(n)$ approaches $V$ as the walker visits all sites for asymptotically large $n$.
These regimes, with the exception of the trivial small $n$-range, can be clearly distinguished in figure \ref{walk}, which shows results for fixed graph size $V=10^4$ and graphs with $\langle k\rangle=4$; plots for other graph sizes and average degrees look qualitatively identical.

\begin{figure}[!ht]
        \centering
            \subfloat[]{ \includegraphics[width=7.cm]{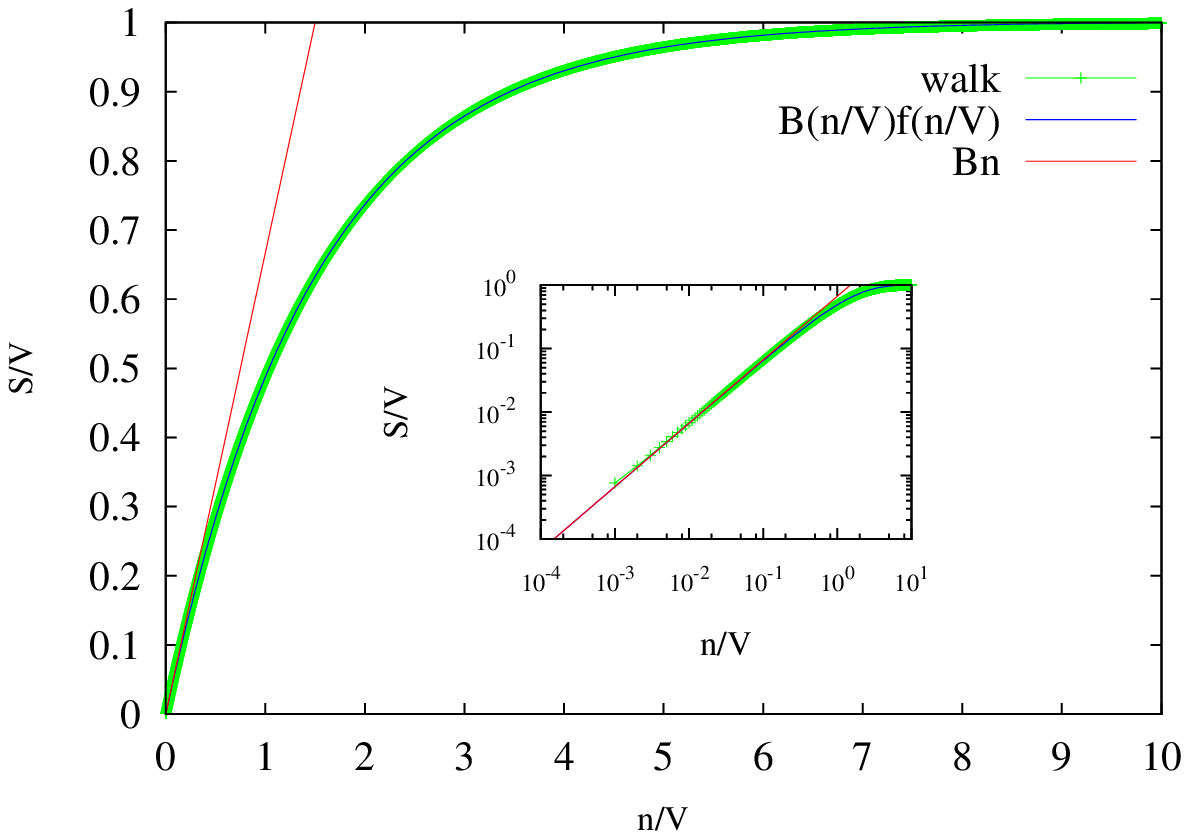}}
               \subfloat[]{ \includegraphics[width=7.cm]{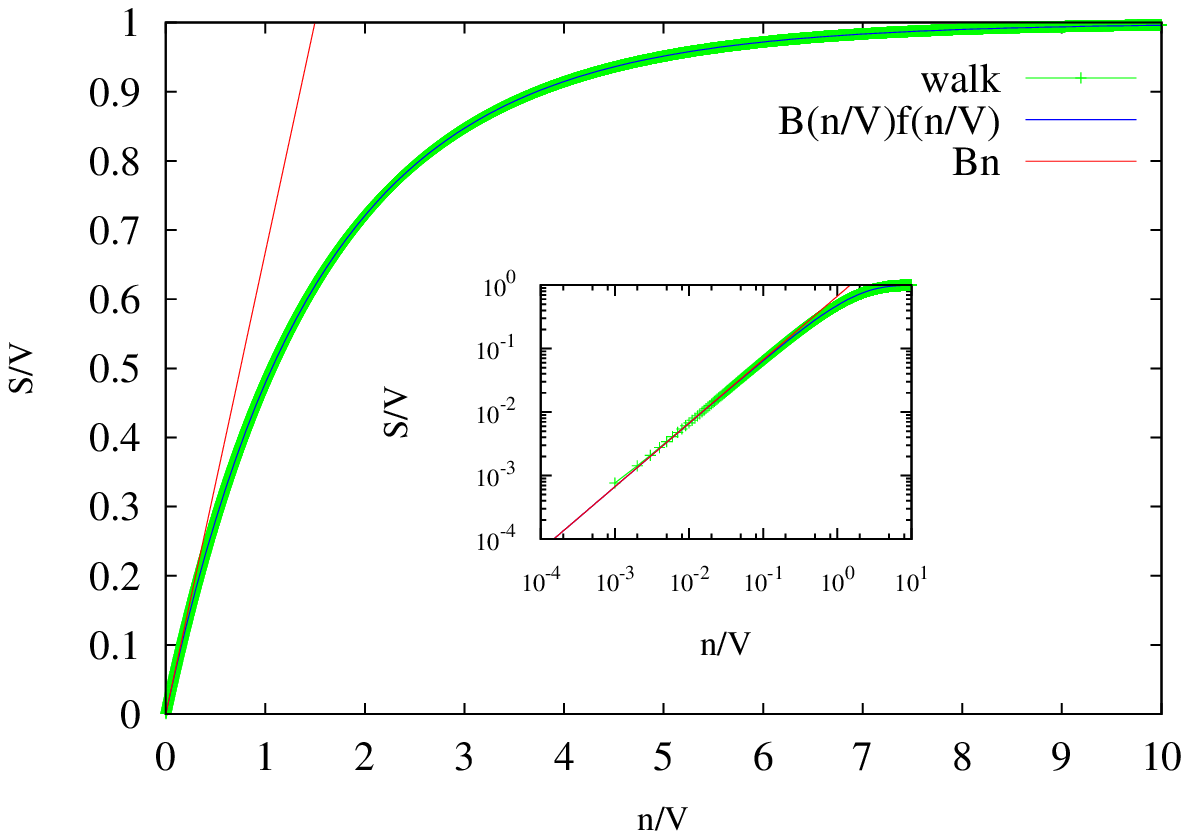}}\\
                 \subfloat[]{ \includegraphics[width=7.cm]{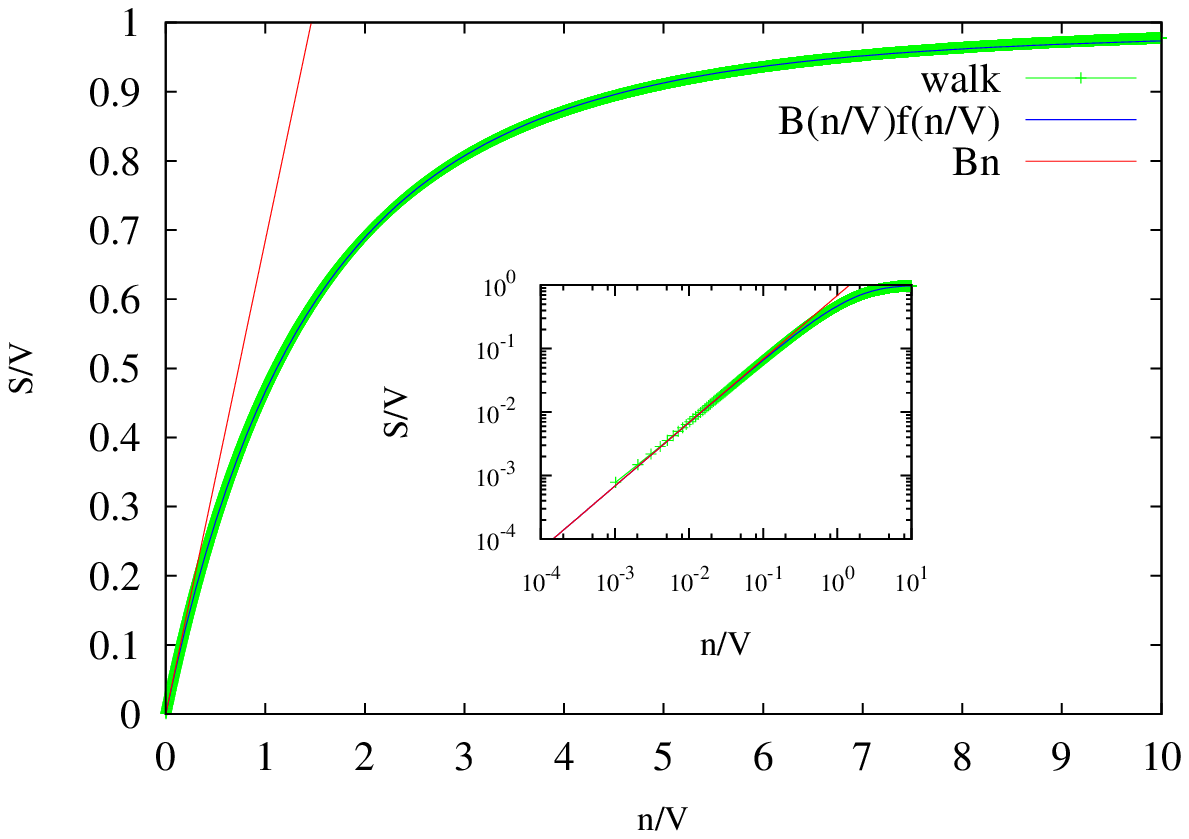}}
                  \subfloat[]{ \includegraphics[width=7.cm]{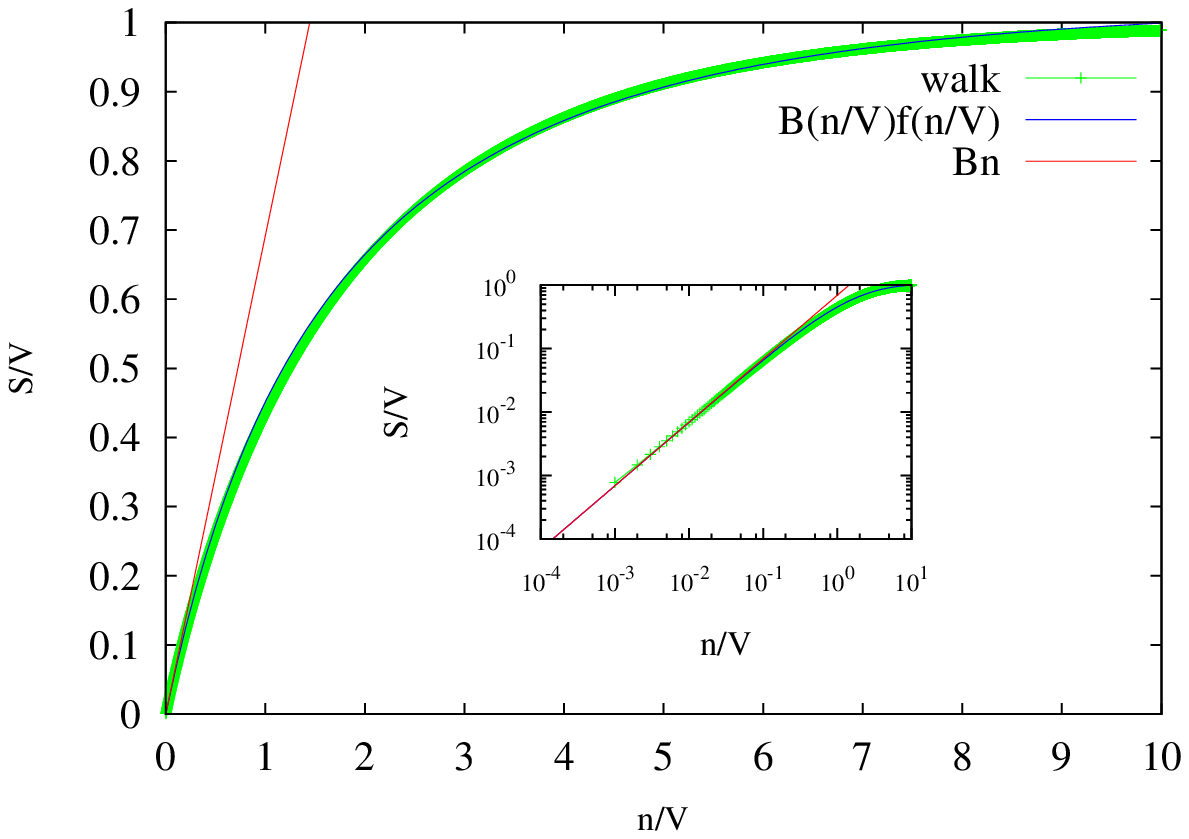}}
                          \caption{Finite size effects: we show the walker behavior by plotting $S(n)/V$, i.e. the fraction of distinct sites visited, derived from direct simulations vs $n/V$. Results are from averages over $1000$ instances of graphs of fixed size $V=10^{4}$ and average degree $\langle k \rangle=4$, for different graph topologies: a) Regular; b) RER; c) ER; d) SF.   The dashed red lines show the cavity predictions $Bn$ for the linear growth with $n$, a regime which is clearer in the log-log plot insets. Beyond that one observes a slow crossover, with $S(n)/V$ eventually approaching unity. Solid lines show our phenomenological scaling fits.}

\label{walk}
\end{figure}

A plausible scaling ansatz that encompasses the various regimes -- again without the initial small $n$-piece -- is
\begin{equation}
\label{scaling_ansatz}
S(n,V) = Bn\, f{\left(\frac{Bn}{V}\right)} 
\end{equation}
where the limiting behaviour of the scaling function must be
\begin{equation} f(x) \approx \left\{
\begin{array}{l l}
1 & \quad  x \ll 1\\
\frac{1}{x} & \quad x \gg 1
\end{array}
\right.
\end{equation}
to reproduce $S(n,V)\approx Bn$ and $S(n,V)\approx V$ when $n$ is much smaller and much larger than $V$, respectively.

In figure \ref{vscaling} we check to what extent the finite-size scaling (\ref{scaling_ansatz}) captures our simulation data. We show results for graph sizes $V=10^{3},10^{4},10^{5}$ and two values for the average degree $\langle k \rangle =4,10$. By plotting $S(n)/(Bn)$ vs $Bn/V$ with $B$ predicted from the cavity equations, we directly have a graphical representation of the scaling function $f(x)$. Very good agreement is seen between the three different graph sizes: these all collapse onto the same curve, except the initial regime discussed above where $S(n)\approx n$ and hence $S(n)/(Bn)>1$. Beyond this we observe a plateau at $S(n)/(Bn)=1$, which in a different guise verifies our claim above that the cavity method does indeed predict the prefactor $B$ correctly. For $x=Bn/V$ growing towards unity, the curves drop below this plateau as expected, indicating the start of the saturation regime. Asymptotically the scaling function $f(x)$ then approaches $1/x$, reflecting the final saturation of $S(n)$ at the upper bound $V$.

More surprising, and not required by our ansatz per se, is that we see in figure \ref{vscaling} good collapse also between graphs of different average degree: using $Bn/V$ as the argument of the scaling function seems sufficient to absorb all the variation with $\langle k\rangle$, without further changes in $f(x)$. The only exception is provided by the scale free graphs, which we discuss in more detail below.

Encouraged by the good agreement of the numerical data with the ansatz (\ref{scaling_ansatz}), we attempt to find simple fits to the scaling function $f(x)$. 
The simulation data show that the crossover starts off with a roughly exponential departure from the small $x$-plateau $f(x)\approx 1$, which suggests a scaling function of the form
$f(x) = {a}/{\ln\!\left( b+(e^{a}-b)e^{ax}\right)}$,
where $a$ and $b$ are fitting parameters. Figure \ref{vscaling} shows that this form fits the data extremely well, and except for the scale-free graphs the fits can be performed even with fixed $b=1$, leaving a single fit parameter.

We comment finally in more detail on the case of SF graphs. Here we see that the data in figure \ref{vscaling} do not collapse perfectly for different $V$ in the intermediate regime where $x=Bn/V$ is order unity or somewhat smaller. In addition, the crossover in $f(x)$ is slower, with $f(x)$ lower in the crossover region than for the other three graph types. We conjecture that both of these effects are due to the presence of many small loops in SF graphs, for example triangles (loops of length 3). To support this hypothesis, we calculated the average number of triangles present in the different types of graph, taking averages over 100 graph instances of size $V=10^{3}$. We found results in the same range for Reg, ER and RER graphs, where the average percentage of nodes that are part of at least one triangle does not exceed $2\%$, $7\%$ and $37\%$ for $\langle k \rangle=4,6,10$ whereas for SF graphs the relevant fractions of nodes reach $9\%$, $24\%$ and $51\%$ for the same average degrees. These results confirm that SF graphs generated via preferential attachment contain a higher number of short loops than the other topologies. In fact it has been shown by spectral arguments \cite{farkas2001} that, even though the fraction of nodes in triangles will tend to zero for $V \rightarrow \infty$, the growth rate of the number of loops of length $l \geq 4$ exceeds all polynomial growth rates, thus these graphs do not become locally treelike for large $V$. Therefore it is somewhat surprising that the cavity predictions for $B$ are quantitatively accurate even for SF graphs.

\begin{figure}[!ht]
        \centering
            \subfloat[]{ \includegraphics[width=7.cm]{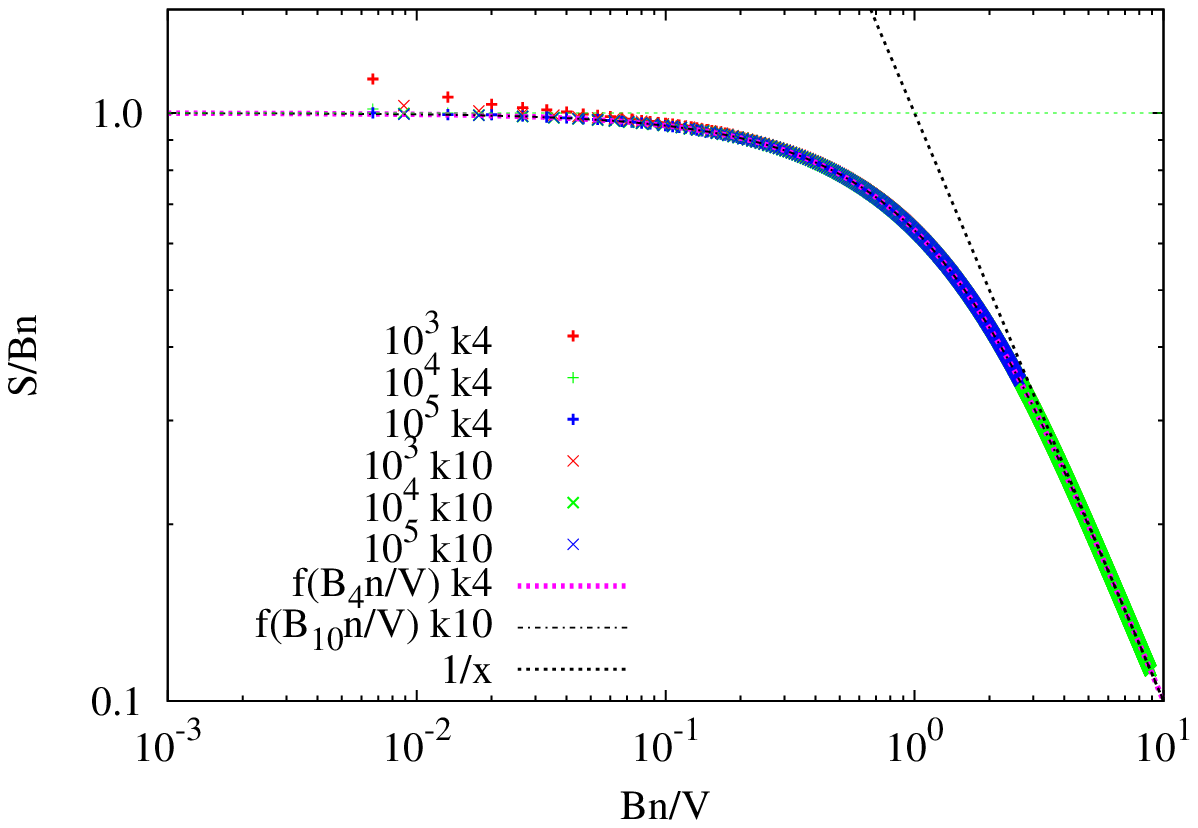}}
               \subfloat[]{ \includegraphics[width=7.cm]{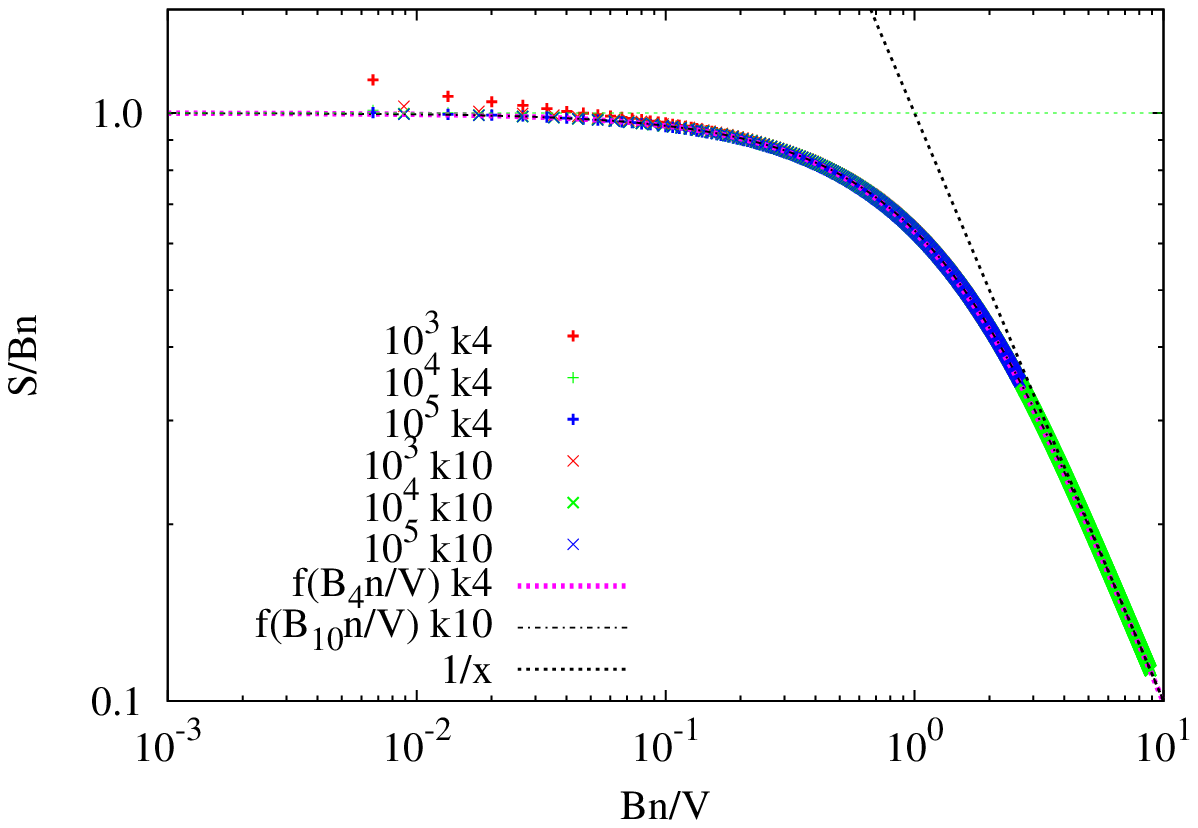}}\\
                 \subfloat[]{ \includegraphics[width=7.cm]{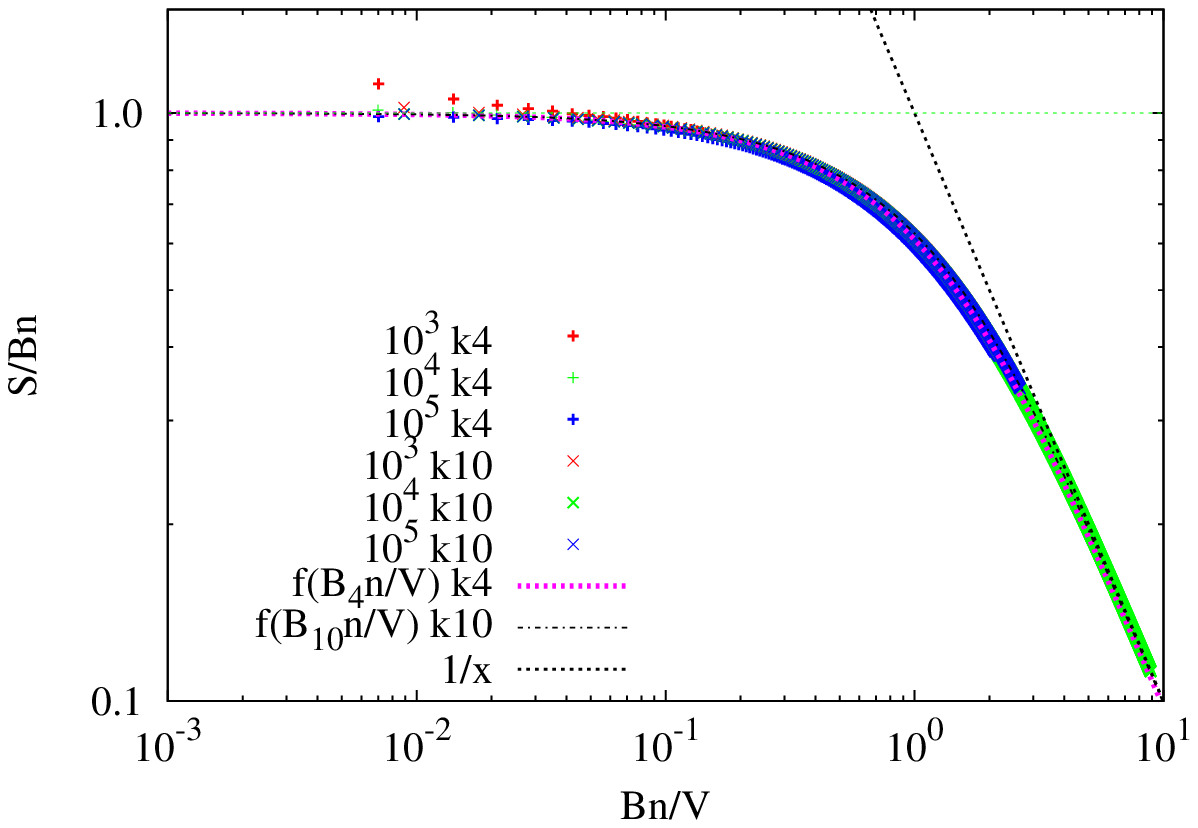}}
                  \subfloat[]{ \includegraphics[width=7.cm]{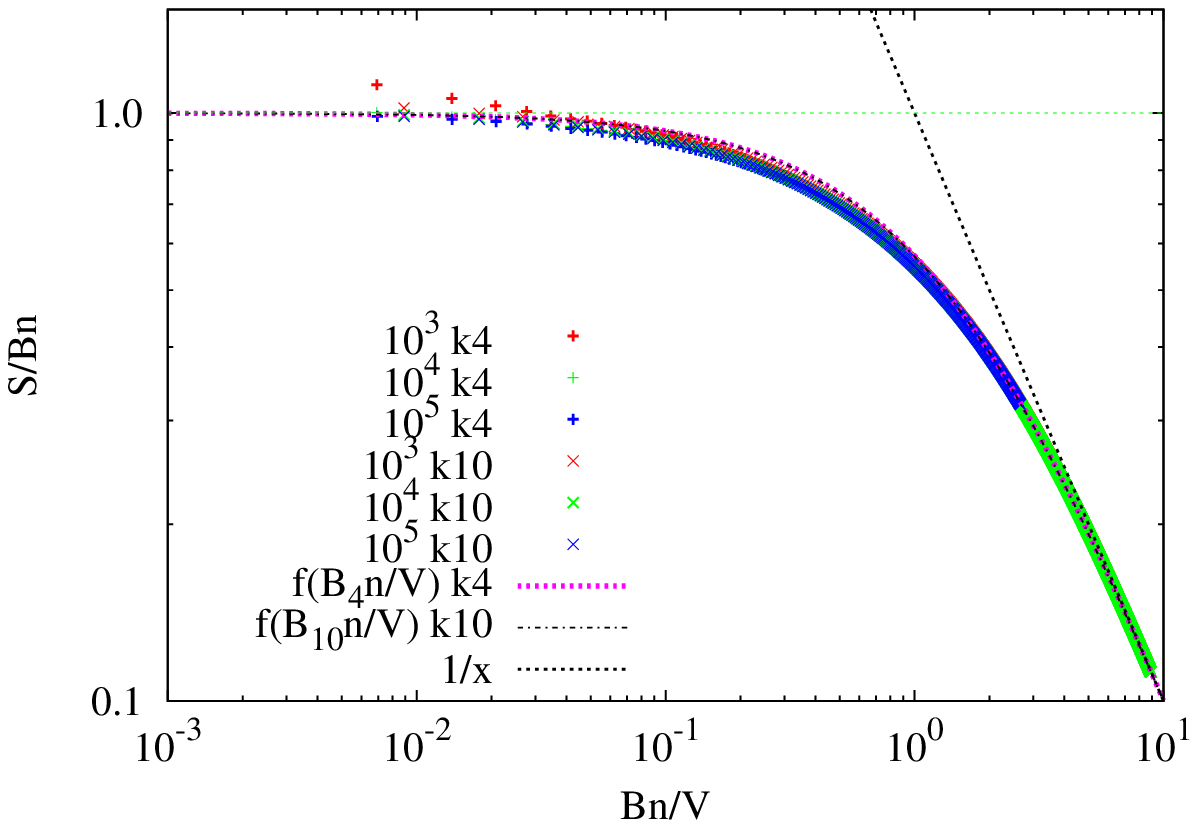}}
                          \caption{Finite-size scaling of number of distinct sites visited, showing $y=S(n)/(Bn)$ versus $x=Bn/V$. Data from direct simulations (symbols), with $B$ predicted from the cavity equations, are shown for graphs of sizes $V=10^{3},10^{4}, 10^{5}$ and average degrees $\langle k \rangle=4,10$.
                          The graph topologies are: (a) Regular; (b) RER; (c) ER; (d) SF. Very good collapse onto a master curve $y=f(x)$ is seen between the different average degrees and -- in (a,b,c) -- also different $V$.
                          The initial plateau at $y=1$ shows the agreement between direct simulations and cavity predictions. For larger $x$ saturation sets in, with $f(x)\approx 1/x$ asyptotically (dotted black line).
                          }\label{vscaling}
\end{figure}

\section{Conclusions.}\label{conclusions}
We have presented an analytical expression for the topology dependent prefactor $B$ governing the linear regime for the average number of distinct sites $S(n)$ visited by a long (large $n$) random walk on a large random graph. We adapted the general results derived for $S(n)$ in terms of generating functions, as used to study $d$-dimensional lattices, to the case of random networks. We then combined message-passing techniques and the properties of Gaussian multivariate distributions to derive an expression for $B$ that is valid for locally tree-like graph structures, and found good agreements between the theoretical predictions and direct numerical simulations. An intriguing feature of the results is that at fixed average degree $\langle k\rangle$, $B$ seems smallest for regular graphs and increases with the width of the degree distribution, and one may conjecture that the regular graph result $B =(k-2)/(k-1)$ is in fact a lower bound.

We analysed finite-size effects for $S(n,V)$ and proposed a simple scaling ansatz to capture these. Apart from a trivial small $n$-regime, one finds a linear regime $S\approx Bn$ with prefactor $B$ in accord with our predictions; an asymptotic regime $Bn\gg V$ where the random walk saturates and $S\to V$; and a crossover in between. Our data provides excellent support for the scaling description, except possibly for SF graphs built via preferential attachment,
and we were able to provide a simple two-parameter (in fact often one-parameter) fit for the scaling function.

The accurate results we obtained using message-passing techniques may open new perspectives in the analysis of random walks on networks. The cavity method we applied
to study random walks on networks could be considered as a valid alternative tool to analyse other types of quantities related to this problem. For instance one could develop further the model by considering a set of $N$ independent random walkers over a random network and studying the behavior of the average number of distinct or common visited sites, as has been done in the case of lattices \cite{Larralde,Majumdar,majumdar2012}. This could give insights into the occupancy statistics of packet-switched networks where packets of data move by independently hopping along nodes to transmit informations between users. The general character of our analysis suggests to us that it should be feasible to adapt it to the study of this or similar types of questions that arise in the study of random walks on networks. 

\section*{Acknowledgement}
This work is supported by the Marie Curie Training Network NETADIS (FP7, grant 290038), the ANR Grant No. 2011-BS04-013-01 WALKMAT and in part by the Indo-French Centre for the Promotion of Advanced Research under Project No. 4604-3. We acknowledge very helpful conversations with Reimer K\"uhn.

\begin{appendices}

\section{The graph representation of the Gaussian covariate distribution. }\label{apxP(x)}
We can rewrite the joint distribution (\ref{multilap}) using $\hat{R}^{-1}_{ij}(z)=\delta_{ij}-z \frac{a_{ij}}{\sqrt{k_{i}k_{j}}}$. In this way we can separate the node and edge contributions respectively to obtain a graphical model representation: 
\begin{eqnarray}
P(\bar{x})&\sim& e^{-\bar{x}^{T}\hat{R}^{-1}(z)\bar{x}/2} \nonumber\\
&=& \exp\left(-\sum_{i}x_{i}[\hat{R}^{-1}(z)\bar{x}]_{i}/2\right) \nonumber\\\
&=& \exp\left(-\sum_{i}x_{i}[\sum_{j}\hat{R}^{-1}_{ij}(z)x_{j}]/2\right) \nonumber\\\
&=&  \exp\left(-\sum_{i}x_{i}[\sum_{j}(\delta_{ij}-z \frac{a_{ij}}{\sqrt{k_{i}k_{j}}})x_{j}]/2\right) \nonumber\\\
&=&  \exp\left(-\sum_{i}x_{i}[x_{i}-z\sum_{j\in \partial i}\frac{x_{j}}{\sqrt{k_{i}k_{j}}}]/2\right) \nonumber\\\
&=&  \exp\left(-\sum\limits_{i}\left\{\frac{1}{2}x_{i}^{2}-\frac{1}{2}zx_{i}\sum\limits_{j\in \partial i}\frac{x_{j}}{\sqrt{k_{i}k_{j}}}\right\}\right) \nonumber\\\
&=& \prod_{i\in \mathcal{V}}e^{-\frac{1}{2}x_{i}^{2}}\prod_{(ij)\in \mathcal{E}}e^{z\frac{x_{i}x_{j}}{\sqrt{k_{i}k_{j}} }} 
\end{eqnarray} 

\section{Regular graph case.}\label{apxrg}
We calculate an exact expression for the topology dependent prefactor in the case of a regular graph. Using $k_{i}=k$, $v_{i}^{(j)}=v^{(j)}$, $m_{i \rightarrow j}=m$, $v_{j}=v$, (\ref{cavmar}) and (\ref{2ndm}) we get:
\begin{eqnarray}
v&=&k\, [k-\sum_{k \in \partial i} m]^{-1} \nonumber \\
&=& k\,[k- k m]^{-1}   \nonumber \\
&=&k\, \left[k\left(1- \frac{1}{k-1}\right)\right]^{-1}  \nonumber \\
&=&k\, \frac{k-1}{k(k-2)}
\end{eqnarray}

 We substitute into the expressions (\ref{BMPv}) for the prefactor $B$ to obtain:
\begin{eqnarray}
B&=&\frac{1}{V  k } \sum_{j \in V} \frac{k}{v} \nonumber  \\
&=& \frac{1}{V  k } \frac{Vk(k-2)}{k-1}  \nonumber  \\
&=&  \frac{k-2}{k-1}
\end{eqnarray}
Therefore the large time limit of the average number of distinct sites of a random walk on a $k$-regular graph is:
\begin{equation}
\lim_{n \rightarrow \infty} S(n) = \left(  \frac{k-2}{k-1} \right) n
\end{equation}

\end{appendices}


\section*{References.}

\bibliography{bibliography}{}

\begin{thebibliography}{10}

\bibitem{target}
Jasch F and Blumen A.
\newblock Target problem on small-world networks.
\newblock {\em Physical Review E}, 63(4):041108, 2001.

\bibitem{trapping}
Jasch F and Blumen A.
\newblock Trapping of random walks on small-world networks.
\newblock {\em Physical review. E, Statistical, nonlinear, and soft matter
  physics}, 64(6 Pt 2):066104--066104, 2001.

\bibitem{beeler}
Beeler~Jr JR.
\newblock Distribution functions for the number of distinct sites visited in a
  random walk on cubic lattices: Relation to defect annealing.
\newblock {\em Physical Review}, 134(5A):A1396, 1964.

\bibitem{relaxation}
Klafter J and Blumen A.
\newblock Models for dynamically controlled relaxation.
\newblock {\em Chemical physics letters}, 119(5):377--382, 1985.

\bibitem{cattuto}
Cattuto C, Barrat A, Baldassarri A, Schehr G, and Loreto V.
\newblock Collective dynamics of social annotation.
\newblock {\em Proceedings of the National Academy of Sciences},
  106(26):10511--10515, 2009.

\bibitem{networking}
Yeung~C H and Saad D.
\newblock Networking—a statistical physics perspective.
\newblock {\em Journal of Physics A: Mathematical and Theoretical},
  46(10):103001, 2013.

\bibitem{Larralde}
Larralde H, Trunfio P, Havlin S, Stanley~H E, and Weiss~G H.
\newblock Number of distinct sites visited by n random walkers.
\newblock {\em Physical Review A}, 45(10):7128, 1992.

\bibitem{Majumdar}
Kundu A, Majumdar~S N, and Schehr G.
\newblock Exact distributions of the number of distinct and common sites
  visited by n independent random walkers.
\newblock {\em Physical review letters}, 110(22):220602, 2013.

\bibitem{vineyard}
Vineyard~G H.
\newblock The number of distinct sites visited in a random walk on a lattice.
\newblock {\em Journal of Mathematical Physics}, 4(9):1191--1193, 1963.

\bibitem{montroll}
Montroll~E W and Weiss~G H.
\newblock Random walks on lattices. ii.
\newblock {\em Journal of Mathematical Physics}, 6(2):167--181, 1965.

\bibitem{erdosdistinct}
Dvoretzky A and Erd{\"o}s P.
\newblock Proceedings of the second berkeley symposium.
\newblock 1951.

\bibitem{RWbethe}
Hughes~Barry D and Sahimi M.
\newblock Random walks on the bethe lattice.
\newblock {\em Journal of Statistical Physics}, 29(4):781--794, 1982.

\bibitem{burioni2005}
Burioni R and Cassi D.
\newblock Random walks on graphs: ideas, techniques and results.
\newblock {\em Journal of Physics A: Mathematical and General}, 38(8):R45,
  2005.

\bibitem{redner2001}
Redner S.
\newblock {\em A guide to first-passage processes}.
\newblock Cambridge University Press, 2001.

\bibitem{WS}
Watts~D J and Strogatz~S H.
\newblock Collective dynamics of ‘small-world’ networks.
\newblock {\em Nature}, 393(6684):440--442, 1998.

\bibitem{SF}
Barab{\'a}si A-L and Albert R.
\newblock Emergence of scaling in random networks.
\newblock {\em Science}, 286(5439):509--512, 1999.

\bibitem{sg87}
M{\'e}zard M, Parisi G, and Virasoro~M A.
\newblock {\em Spin glass theory and beyond}, volume~9.
\newblock World scientific Singapore, 1987.

\bibitem{rogers2008}
Rogers T, Castillo~I P, K{\"u}hn R, and Takeda K.
\newblock Cavity approach to the spectral density of sparse symmetric random
  matrices.
\newblock {\em Physical Review E}, 78(3):031116, 2008.

\bibitem{ipc}
M{\'e}zard M and Montanari A.
\newblock {\em Information, physics, and computation}.
\newblock Oxford University Press, 2009.

\bibitem{noh}
Noh~J D and Rieger H.
\newblock Random walks on complex networks.
\newblock {\em Physical review letters}, 92(11):118701, 2004.

\bibitem{fisher1984}
Fisher~M E.
\newblock Walks, walls, wetting, and melting.
\newblock {\em Journal of Statistical Physics}, 34(5-6):667--729, 1984.

\bibitem{spectral}
Chung~F RK.
\newblock {\em Spectral graph theory}, volume~92.
\newblock American Mathematical Soc., 1997.

\bibitem{braket}
Dirac P~A M.
\newblock A new notation for quantum mechanics.
\newblock {\em Mathematical Proceedings of the Cambridge Philosophical
  Society}, 35:416--418, 7 1939.

\bibitem{frobenius}
Frobenius G.
\newblock Über matrizen aus nicht negativen elementen.
\newblock {\em Sitzungsber. Königl. Preuss. Akad. Wiss. Berlin}, page
  456–477, 1912.

\bibitem{perron}
Perron O.
\newblock Zur theorie der matrices.
\newblock {\em Mathematische Annalen}, 64(2):248--263, 1907.

\bibitem{friedman1991}
Friedman J.
\newblock On the second eigenvalue and random walks in random d-regular graphs.
\newblock {\em Combinatorica}, 11(4):331--362, 1991.

\bibitem{broder1987}
Broder A and Shamir E.
\newblock On the second eigenvalue of random regular graphs.
\newblock In {\em Foundations of Computer Science, 1987., 28th Annual Symposium
  on}, pages 286--294. IEEE, 1987.

\bibitem{farkas2001}
Farkas~I J, Der{\'e}nyi I, A-L Barab{\'a}si, and Vicsek T.
\newblock Spectra of “real-world” graphs: Beyond the semicircle law.
\newblock {\em Physical Review E}, 64(2):026704, 2001.

\bibitem{furedi81}
Füredi Z and Komlós J.
\newblock The eigenvalues of random symmetric matrices.
\newblock {\em Combinatorica}, 1(3):233--241, 1981.

\bibitem{chung2003}
Chung F, Lu~L, and Vu~V.
\newblock Spectra of random graphs with given expected degrees.
\newblock {\em Proceedings of the National Academy of Sciences},
  100(11):6313--6318, 2003.

\bibitem{edwards1976}
Edwards~S F and Jones~R C.
\newblock The eigenvalue spectrum of a large symmetric random matrix.
\newblock {\em Journal of Physics A: Mathematical and General}, 9(10):1595,
  1976.

\bibitem{wormald1999}
Wormald~N C.
\newblock Models of random regular graphs.
\newblock {\em London Mathematical Society Lecture Note Series}, pages
  239--298, 1999.

\bibitem{ER}
Erd\H{o}s P and R\'{e}nyi A.
\newblock On the evolution of random graphs.
\newblock {\em Publications of the Mathematical Institute of the Hungarian
  Academy of Sciences}, 5:17--61, 1960.

\bibitem{Monasson}
Monasson R.
\newblock Diffusion, localization and dispersion relations on “small-world”
  lattices.
\newblock {\em The European Physical Journal B-Condensed Matter and Complex
  Systems}, 12(4):555--567, 1999.

\bibitem{peter2013}
Urry~M J and Sollich P.
\newblock Random walk kernels and learning curves for {Gaussian} process
  regression on random graphs.
\newblock {\em The Journal of Machine Learning Research}, 14(1):1801--1835,
  2013.

\bibitem{peter2012}
Urry~M J and Sollich P.
\newblock Replica theory for learning curves for {Gaussian} processes on random
  graphs.
\newblock {\em Journal of Physics A: Mathematical and Theoretical},
  45(42):425005, 2012.

\bibitem{sollich2014}
Sollich P, Tantari D, Annibale A, and Barra A.
\newblock Extensive load in multitasking associative networks.
\newblock {\em arXiv:1404.3654}, 2014.

\bibitem{reimer2008}
K{\"u}hn R.
\newblock Spectra of sparse random matrices.
\newblock {\em Journal of Physics A: Mathematical and Theoretical},
  41(29):295002, 2008.

\bibitem{alon1986}
Alon N.
\newblock Eigenvalues and expanders.
\newblock {\em Combinatorica}, 6(2):83--96, 1986.

\bibitem{nilli1991}
Nilli A.
\newblock On the second eigenvalue of a graph.
\newblock {\em Discrete Mathematics}, 91(2):207--210, 1991.

\bibitem{majumdar2012}
Majumdar~S N and Tamm~M V.
\newblock Number of common sites visited by n random walkers.
\newblock {\em Physical Review E}, 86(2):021135, 2012.

\end{thebibliography}
\bibliographystyle{unsrt}

\end{document}